\documentclass[aps,prb,10pt,twocolumn]{revtex4}

\usepackage[cmex10]{amsmath}
\usepackage{amsfonts}
\usepackage{amsmath}
\usepackage{amssymb}
\usepackage{graphicx}
\usepackage{multirow}
\usepackage{bm}
\usepackage{color}

\begin{document}
\title{Upmost efficiency, few-micron-sized midwave infrared HgCdTe photodetectors}

\author{Roy Avrahamy}
\affiliation{Department of Electrical and Computer Engineering, Ben-Gurion University of the Negev, P.O.B 653, Beer-Sheva, Israel 84105}
\author{Moshe Zohar}
\affiliation{Department of Electrical and Electronics Engineering, Shamoon College of Engineering, P.O.B. 950, Beer Sheva, Israel 84100}
\author{Mark Auslender}
\affiliation{Department of Electrical and Computer Engineering, Ben-Gurion University of the Negev, P.O.B 653, Beer-Sheva, Israel 84105}
\author{Zeev Fradkin}
\affiliation{Department of Electrical and Electronics Engineering, Shamoon College of Engineering, P.O.B. 950, Beer Sheva, Israel 84100}
\author{Benny Milgrom}
\affiliation{School of Electrical Engineering, The Jerusalem College of Technology, P.O.B 16031, Jerusalem, Israel 91160}
\author{Rafi Shikler}
\affiliation{Department of Electrical and Computer Engineering, Ben-Gurion University of the Negev, P.O.B 653, Beer-Sheva, Israel 84105}
\author{Shlomo Hava}
\affiliation{Department of Electrical and Computer Engineering, Ben-Gurion University of the Negev, P.O.B 653, Beer-Sheva, Israel 84105}

\begin{abstract}
Resonant cavity-assisted enhancement of optical  absorption was a photodetector designing concept emerged about two and half decades ago, which responded to the challenge of thinning the photoactive layer while outperforming the efficiency of the monolithic photodetector. However, for many relevant materials, meeting that challenge with such a design requires unrealistically many layer deposition steps, so that the efficiency at goal hardly becomes attainable because of inevitable fabrication faults. Under this circumstance, we suggest a new approach for designing photodetectors with absorber layer as thin as that in respective resonant cavity enhanced ones, but concurrently, the overall detector thickness being much thinner, and topmost performing. The proposed structures also contain the cavity-absorber arrangement but enclose the cavity by two dielectric one-dimensional grating-on-layer structures with the same grating pitch, instead of the distributed Bragg reflectors typical of the resonant cavity enhancement approach. By design based on the in-house software, the theoretical feasibility of such $\sim 7.0\mu{\rm m}-8.5\mu\rm m$ thick structures with $\sim 100\%$ efficiency for a linearly polarized (TE or TM) mid-infrared range radiation is demonstrated. Moreover, the tolerances of the designed structures' performance against the gratings' fabrication errors are tested, and fair manufacturing tolerance while still maintaining high peak efficiency along with a small deviation of its spectral position off initially predefined central-design wavelength is proved. In addition, the electromagnetic fields amplitudes and Poynting verctor over the cavity-absorber area are visualized. As a result, it is inferred that the electromagnetic fields' confinement in the designed structure, which is a key to their upmost efficiency, is two-dimensional combining in-depth vertical resonant-cavity like confinement, with the lateral microcavity like one set by the presence of gratings.
\end{abstract}

\maketitle

\section{Introduction}
Photodetector (PD) is an indispensable part in optical sensors, communication systems and photonic interconnects. If each photon absorbed in a photoactive layer of PD contributes to creation of an electron–-hole pair and there is no parasitic recombination, the quantum efficiency $\eta$, i.e. the ratio of the number of the photogenerated electron–hole pairs to the number of incident photons, equals the optical absorbance $A$. In monolithic PDs the detected light traverses the absorber once, so $\eta$ may be increased only by increasing the absorber's thickness $t_{\rm a}$, but the larger the $t_{\rm a}$ the stronger the thermal generation–-recombination noise and the longer the transit time in photodiode PDs.

For most applications, a high signal-to-noise ratio and operation speed concurrently with high $\eta$ are both crucial. A drastic decrease in $t_{\rm a}$ to this end while keeping high $\eta$ requires the impinging light to be strongly confined in a vicinity of the absorber. An adopted approach to the confinement is the solid-state resonant Fabry-Perot cavity in which the absorber layer is sandwiched between two multilayers. In each multilayer, a distributed Bragg reflector (DBR) acts as a mirror, and two layers adjacent to the absorber layer form a cavity. Thin enough absorber may weakly affect the optical cavity-resonance condition absorbing nevertheless at the resonance wavelength the largest portion of the incident light. Such a resonant cavity enhanced (RCE) PD \cite{Unlu,{RCE}} enables much more efficient and faster (for photodiode PD) detection as well as operation at higher temperatures with a much thinner absorber than its monolithic counterpart, although these benefits are traded off against a smaller spectral bandwidth.

The field of RCE PDs have matured over the past two and half decades, regarding different spectral ranges, e.g. the telecom bands in the near infrared (NIR) \cite{Wang,{Tuttle},{Alija}}, and mid-wave infrared (MWIR) ranges \cite{Sioma,RCE2,RCEPbX}, as well as suitable physical devices, such as p-i-n, Shottky and avalanche photodiodes \cite{Unlu,RCE,Wang,Tuttle,Alija,RCE3,RCE1},  and photoconductor \cite{Sioma,RCE2,RCEPbX}. Yet, due to restrictions on the mirrors' optical properties \cite{Unlu,RCE}, the RCE PDs designed for topmost $\eta$ should contain large number of layers in their DBRs that it would certainly exceed a technology defined critical value, see e.g. \cite{DBR-MCT} for HgCdTe based PDs, beyond which the DBRs' fabrication inevitably introduces growth defects. In turn, the defects impair the DBRs' reflectance, making attainment $\eta$ at goal impractical.

In a quest for thinning PDs, we proposed to replace the front DBR mirror by an one-dimensional (1D) grating-on-layer structure (GLS), and reported the proof-of-concept examples of HgCdTe based PDs \cite{Zohar,{OQE}}.
The round-trip phase control \cite{Unlu} extended to the presence of GLS and optimization allowed us \cite{Zohar,OQE} to achieve tolerant designs with the $\eta\approx 100\%$ and backside DBR twice as thin as compared to those of an optimized conventional RCE PD comprising the same materials. Still, if the back mirror remains to be a DBR one the RCE device can not be downscaled anymore due to lower bounds imposed on the cavity and DBRs thicknesses by tailoring the round-trip phase and DBR mirrors' reflectance, respectively \cite{Unlu,RCE}.

An emergent promising approach to high capability photo-detection is nowadays with PD focal plane array (FPA) \cite{RogalskiFaraone,Rogalski}. As the monolithic PDs are low efficient, integrating them in FPA requires high-density packaging which may prove faulty. Thus the high capability PD FPAs my benefit integrating upmost efficient PDs, while thinning the latter would be favorable for the planar on-chip applications.

Here, to meet challenge of further reducing the size and complexity of the entire PD device by surpassing the above thickness bounds while benefiting from our previous designs \cite{Zohar,{OQE}} which attained uppermost $\eta$, we propose to replace the back DBR mirror in those designs also by an 1D GLS with the same grating period as that in the front one. High refractive index (RI) contrast GLSs are attracting much interest \cite{Chang} as a new platform for integrated optoelectronics due to their ability to function in some spectral range quite {\em alike} usual optical elements, e.g. mirrors, being much \emph{thinner}.
\vspace{-1mm}
\section{Proposed photodetector structures}
We consider PD structures, adapted to MWIR range portion, around a center-design wavelength (CDW), using HgCdTe, as the absorber. The MWIR design involves a variety of materials, see \cite{Sioma,RCE2,DBR-MCT,Faraone2} and below. Normal-incidence backside (through-substrate) illumination (BSI) is used all over the study. Benefiting the absence of electrical contacts at the PD backside, BSI has an essential advantage of reasonable reducing the PD pixel without decreasing the amount of the input light power \cite{RCE2}.

\begin{figure*}[!ht]
\begin{center}
\includegraphics[width=0.85\textwidth]{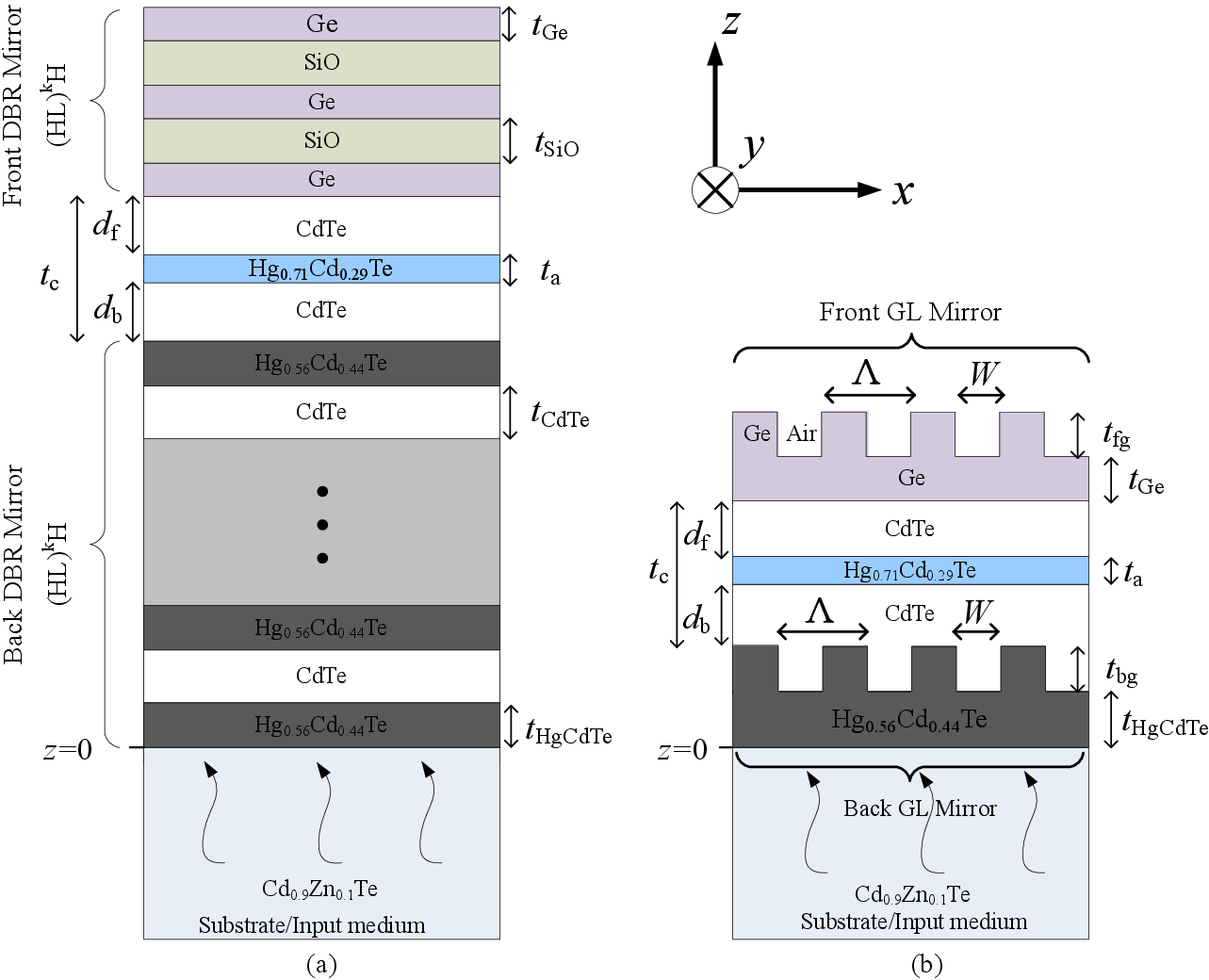}
\caption{{\label{fig:fig1}} The MWIR PDs structures, composed with an Hg$_{0.71}$Cd$_{0.29}$Te absorber layer in a cavity and: (a) two DBR mirrors (RCE PD); (b) two mirror-like GLSs (dual-GLS PD). In the coordinate system above the dual-GLS PD's schematic, the grating's periodicity and grooves/lines are along the $x$- and $y$-axes, respectively; the layers' stacking and light impinging directions are along the $z$-axis.}
\end{center}
\end{figure*}

Fig.\ref{fig:fig1} shows the structure schematic, in which Fig.\ref{fig:fig1}(a) depicts MWIR RCE PD of conventional schematic \cite{Unlu,RCE} served us as a reference, and Fig.\ref{fig:fig1}(b) shows its double-GLS modification. The front and back DBRs in Fig.\ref{fig:fig1}(a) are composed of Ge/SiO and Hg$_{0.56}$Cd$_{0.44}$Te/CdTe quarter-wave stacks in respect; CdTe builds the cavity ambient of the Hg$_{0.71}$Cd$_{0.29}$Te layer, while the substrate is Cd$_{0.9}$Zn$_{0.1}$Te \cite{RCE2,DBR-MCT,Faraone2}. In Fig.\ref{fig:fig1}(b), the absorber, its ambient, the substrate and irradiation mode are the same as in Fig.\ref{fig:fig1}(a), but now the reflectors are: the air-Ge/Ge and CdTe-Hg$_{0.56}$Cd$_{0.44}$Te/Hg$_{0.56}$Cd$_{0.44}$Te GLSs.
%Fig 1

\section{Statement of problem and principles of modeling}
In this study, we adopt the $\eta=A$ modeling that simplifies the design to one in which only the optics needs to be considered, and still has a prevalent use for initial PD designs \cite{Unlu,RCE,RCE1,RCE2,Zohar}. An electronic-device simulation can be done, see e.g. \cite{RCE1,Faraone2,Medhat},  to complement and refine the optical approach but hardly can disprove its main outcomes. Therefore, rather than being restricted to a specific PD technology, we analyze the structures optically, characterizing the constitutive layers, including the absorber, by complex RIs $n+{\rm i}k$, $n$ and $k\ll n$ being real RIs and extinction coefficients, respectively. Complex RIs of the involved materials, see e.g. \cite{Palik_MCT,Palik_CdTe,Palik_Ge,Palik_SiO,CdZnTe}, at CDW are exposed in Table \ref{tab:t1}.
%------------------------------------------
%Table 1
\begin{table*}[!htb]
\caption{\bf RIs of materials for MWIR PDs at CDW set in the text and shown below.}
\label{tab:t1}
\begin{center}
\begin{tabular}{|c|c|c|c|c|c|c|}
\hline
\rule[-1ex]{0pt}{3.5ex} $\lambda_0,\hspace{0.02cm}\mu {\rm m}$ & Ge   & SiO     & CdTe  & Hg$_{0.56}$Cd$_{0.44}$Te  & Cd$_{0.9}$Zn$_{0.1}$Te & Hg$_{0.71}$Cd$_{0.29}$Te   \\%09.2017
\hline
\rule[-1ex]{0pt}{3.5ex} $4.415$ & $3.9332+{\rm i}\;0$     & $1.78+{\rm i}\;0$    & $2.6695+{\rm i}\;0$ & $2.9709+{\rm i}\;0$ & $2.6896+{\rm i}\;0$  & $3.4826+ {\rm i}\;0.1477$   \\%09.2017
\hline
\end{tabular}
\end{center}
\end{table*}
Neither the width nor shape of the $\eta(\lambda)$ spectra are of our concern, rather we put the emphasis on obtaining maximal CDW efficiency $\eta_{\max}=\max\eta\left(\lambda_0\right)$, while keeping $t_{\rm a}$ constant per a design and the back-grating cladding's thickness, similarly to the DBR layers' ones, fixed at the respective quarter-wavelength value throughout. Such dimensions as the grating pitch $\Lambda$ and groove width $W$ (constrained to be the same for both GLSs), etch depths $t_{\rm fg}$ and $t_{\rm bg}$ as well as the thicknesses of the front-grating cladding and front and back cavities, $d_{\rm f}$ and $d_{\rm b}$ in respect, are the design variables.

It also worthwhile to prevent losses due to light escape from the PD's sides, which is achieved by suppressing non-specular orders of diffraction from GLSs into the cavity. To this end, it is sufficient to consider the subwavelength grating (SWG) operation regime, viz. to constrain the minimum operation wavelength by $\lambda_{\min}>\lambda_{\rm R}$, where $\lambda_{\rm R}=\Lambda n_{\rm CdTe }$. Yet, the SWG regime is not compulsory for the non-specular orders can be suppressed with additional design optimization constraints.

\section{Design procedure}
We performed the designs with a trial-and-error automated multi-start optimization using in-house software, described in detail elsewhere \cite{Zohar}. For trial $d_{\rm f}$ and $d_{\rm b}$ we use values compatible with the cavity thickness assessed from the round-trip phase condition \cite{Unlu}, with the mirrors' reflection phases computed for GLSs at standalone configurations.
While the planar-interface multilayers are insensitive to the polarization of normally incident light \cite{Heavens}, the 1D gratings are strongly sensitive to it. Hence, given $t_{\rm a}$ any our design yields at least three structures -- conventional quarter-wavelength DBR RCE PD, and at least two different dual-GLS based PDs for TE- and TM-polarized irradiation, labeled further as 2DBR, and 2GLS-TE and 2GLS-TM, in respect.
When optimizing RCE PD, we kept it comparable in thickness to corresponding thickest optimized dual-GLS PD.

\section{Structural parameters and efficiency spectra}
%Structural parameters
%------------------------------------------------------
%MCT Table 2
\begin{table*}[!htb]
\centering
\caption{{\bf The parameters of the MWIR PD structures optimally designed with $t_{\rm{a}}=0.075\hspace{0.02cm}\mu \rm m$ as set in the text}$^\ast$}
\vspace*{2mm}
\label{tab:t2}
\begin{tabular}{|c|c|c|c|c|c|c|c|c|c|c|c|}
\hline
 \rule[-1ex]{0pt}{3ex}Structure & $t_{\rm{bg}}$ & $d_{\rm{b}}$  & $d_{\rm{f}}$  & $t_{\rm{Ge}}$ & $t_{\rm{fg}}$ & $\Lambda$ & $W$   & $\eta_{\max}$ & $\lambda_{\rm{R}}$ & $N$ & $t$\\
 \hline
 %\multicolumn{12}{|c|}{$t_{\rm{a}}=0.075$ } \\
% \hline
  2DBR            &        --     & 0.271         & 0.435         & --                 & --            & --        & --     & 0.502                    & --             &23& 8.733 \\
\hline
  2GLS-TE         &        0.600  & 0.000         & 4.075         & 1.005              & 0.888         & 1.594     & 0.836  & 0.999                    & 4.255          &6 &7.015 \\
\hline
  2GLS-TM         &        0.785  & 4.803         & 0.358         & 2.000              & 0.096         & 1.600     & 0.701  & 0.995                    & 4.271          &7 &8.487 \\
\hline
\end{tabular}
\begin{flushleft} $^\ast$All dimensional parameters above are in microns. $t$ is the thickness of the structure excluding the substrate, $N$ is the total number of layers (including gratings, if any), $\eta_{\max}$ is the peak efficiency, $\lambda_{\rm R}$ is the Rayleigh cutoff discussed in the text, and other dimensions are described there and shown in Fig.\ref{fig:fig1}.
\end{flushleft}
\end{table*}
%Fig 2
\begin{figure}[!htb]
\begin{center}
\includegraphics[width=0.47\textwidth]{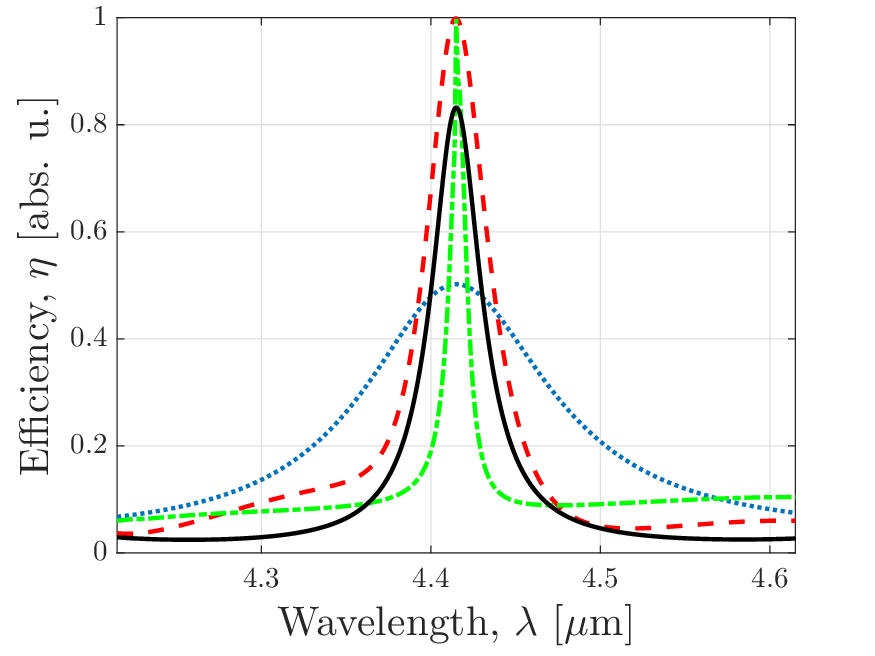}
\caption{{\label{fig:fig2}} The efficiency spectra of the MWIR PDs, designed as described in the text, with the 0.075 $\mu \rm{m}$ thick Hg$_{0.71}$Cd$_{0.29}$Te absorber layer. Broken lines: red -- 2GLS-TE, green --  2GLS-TM, blue -- 2DBR; full black line: 2DBR$_{\rm tu}$. The parameters of the structures are found in Table \ref{tab:t2}.}
\end{center}
\end{figure}

We performed design and simulation with the absorber thickness $t_{\rm a}=0.75\hspace{0.02cm}\mu\rm m$ well tested in trials of the RCE HgCdTe PDs fabrication \cite{RCE2,Faraone2}. The optimization of dual-GLS PD for the TE polarization drives to a structure with $d_{\rm{b}}=0$, i.e. that in which the absorber is adjacent to the back grating's top, see the parameters of the 2GLS-TE structure in Table \ref{tab:t2}. For the structures shown in Fig.\ref{fig:fig1}, the optimal design results including $\eta_{\max}$ are shown in Table \ref{tab:t2}, whereas the $\eta(\lambda)$ spectra are shown in Fig.\ref{fig:fig2}.

The design of reference RCE PD under the above thickness constraint results in the 2DBR structure with the (Ge/SiO)$^{\rm k}$Ge front and (Hg$_{0.56}$Cd$_{0.44}$Te/CdTe)$^{\rm m}$Hg$_{0.56}$Cd$_{0.44}$Te back DBR patterns, see in Fig.\ref{fig:fig1}(a), where ${\rm k}=2$, ${\rm m}=7$, so $N=23$ as shown in Table \ref{tab:t2}. The data for the 2DBR structure in Table \ref{tab:t2} show that it meets the relation $t_{\rm a}\ll d_{\rm f}+d_{\rm b}$ consistent with the canonical RCE approach \cite{Unlu,RCE}, though the absorber is drastically offset from the cavity center. Releasing the above thickness constraint and searching instead RCE PD with $\eta_{\max}$ similar to that of obtained dual-GLS PDs, results in $N=51$ ($\eta_{\max}= 99.9\%$), which is unfeasible with the contemporary HgCdTe technology \cite{RCE2,Faraone2}. Yet, at technologically utmost $N=39$ we obtain optimal RCE PD, notated \emph{ad hoc} 2DBR$_{\rm tu}$, with $\eta_{\max}=83.1\%$ and $t=15.01\mu \rm{m}$ \cite{Zohar}, which is $\sim 1.7$ times thicker than 2DBR in Table \ref{tab:t2}.

As seen from Table \ref{tab:t2} and Fig.\ref{fig:fig2}, designed dual-GLS PDs highly outperform optimal RCE PD with comparable thickness -- the 2DBR structure, the efficiency spectrum of which is the widest one among the spectra shown in Fig.\ref{fig:fig2}. In turn, 2DBR PD outperforms any monolithic HgCdTe based PD with similar $t_{\rm a}$. 2GLS-TE and 2GLS-TM PDs exhibit the widest and narrowest $\eta(\lambda)$ spectra, respectively, whereas FWHM of the 2DBR$_{\rm tu}$ PD efficiency spectrum lies in between corresponding FWHMs for 2GLS-TE and 2GLS-TM PDs, see in Fig.\ref{fig:fig2}.

\section{Electromagnetic confinement properties}
As discussed in the Introduction, conventional planar RCE PD realizes the vertical (1D) light confinement due to the optical resonance \cite{Unlu,RCE}. For the thinnest device, the confinement scales with an effective wavelength in the cavity-absorber region. Since dual-GLS PDs under study are assemblies of identical vertical ($z$) top-to-substrate microscale cuts of the width $\Lambda$ from the structures shown in Fig.\ref{fig:fig1}(b), another confinement physics, which is electromagnetic (EM) field resonant recirculation in micro/nanoscale areas or volumes on sub-wavelength scales \cite{Vahala}, might be more suitable for these structures. We will pursue viewing the above cut as ``microcavity'' (quotes indicate that it has no real vertical boundaries). There were studies which put forward the nano/microcavity analogy for another grating based optical resonator \cite{Fainman} and two-dimensional (2D) photonic crystal slabs \cite{Lalanne}.

While the phenomenon behind the resonant-cavity EM confinement is well understood, the current understanding of that in micro/nanocavity structures, which would be more structure-specific, is nowadays much less mature. Computation of the electric and magnetic fields ($\mathbf{E}$ and $\mathbf{H}$), and Poynting vector ($\mathbf{S}$) at CDW and visualizing its results, are to shed light and infer on the near-EM field recirculation in dual-GLS PDs, while for RCE PDs aim primarily at confirming the adopted vision \cite{Unlu,RCE}. Visualization of $\mathbf{S}$ is expedient for inferring on the EM power flow. We restrict the graphic presentation to $z$ within the absorber-cavity region.

\begin{figure}[!htb]
\begin{center}
\includegraphics[width=0.47\textwidth]{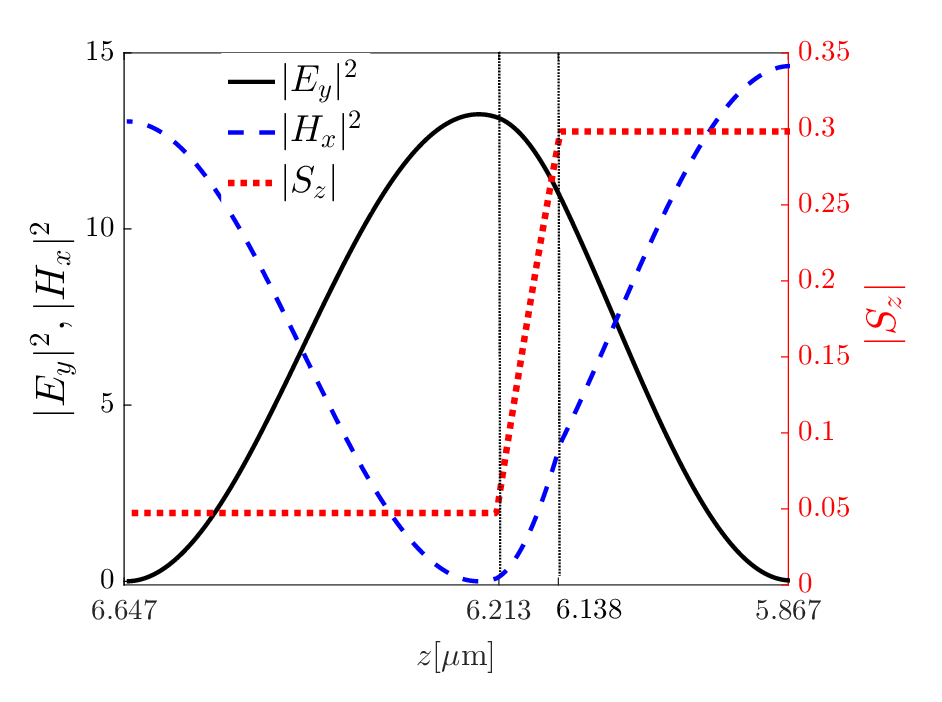}
\caption{{\label{fig:fig3}} The electric and magnetic fields' amplitudes squared (left ordinate) and the Poyinting vector amplitude (right ordinate), normalized to those in the input medium, across the cavity-absorber region of the RCE MWIR 2DBR structure designed as described in the text, the parameters of which are found in Table \ref{tab:t2}. Simulated at CDW $\lambda_0=4.415 \mu m$.}
\end{center}
\end{figure}

For RCE PD, the only nonzero components $E_y$, $H_x$ and $S_z$ are independent of $x$, so we plotted the graphs of $\left|E_y\right|^2$, $\left|H_x\right|^2$ and $\left|S_z\right|$ vs. $z$, shown in Fig.\ref{fig:fig3} where the intersections of the dashed vertical line with the abscissa mark the absorber range. As seen, the fields's amplitudes squared are indeed concentrated in the cavity-absorber region showing there large relative-to-input values, and $\left|S_z\right|$ is constant in the cavity of the RCE structure that may be expected \cite{Unlu,RCE}. $\max\limits_{z}\left|E_y\right|^2\sim \max\limits_{z}\left|H_x\right|^2$ and $\left|S_z\right|$ is fairly linear in $z$ across the absorber.

%Fig 3

\begin{figure*}[!htb]
\includegraphics[width=1.0\textwidth]{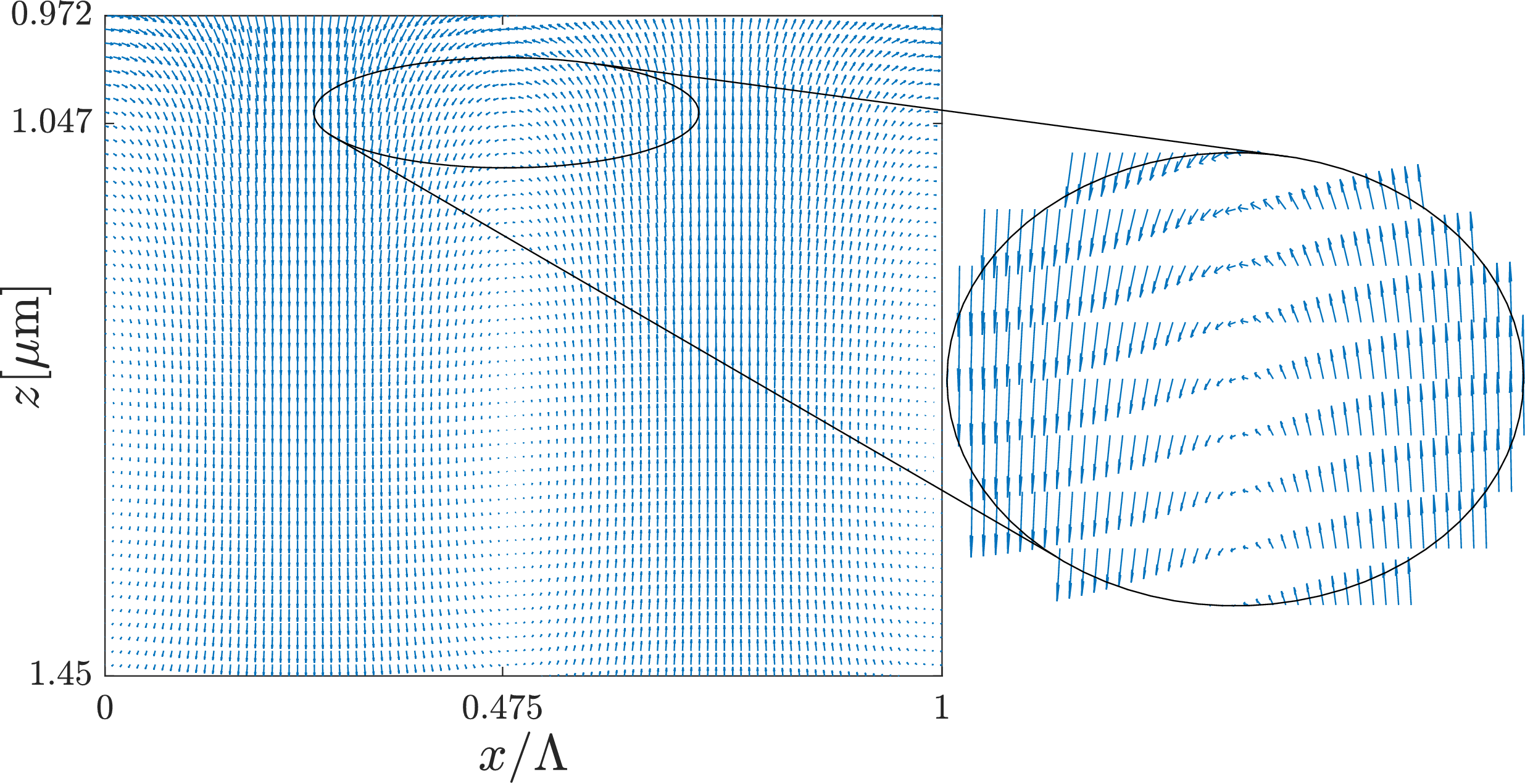}
\caption{{\label{fig:fig4}} The EM power flow map in the structure 2GLS-TE (Table \ref{tab:t2}). The quiver plot range and simulation wavelength are the same as in Fig.\ref{fig:fig6}.}
\end{figure*}

\begin{figure*}[!htb]
\includegraphics[width=1.0\textwidth]{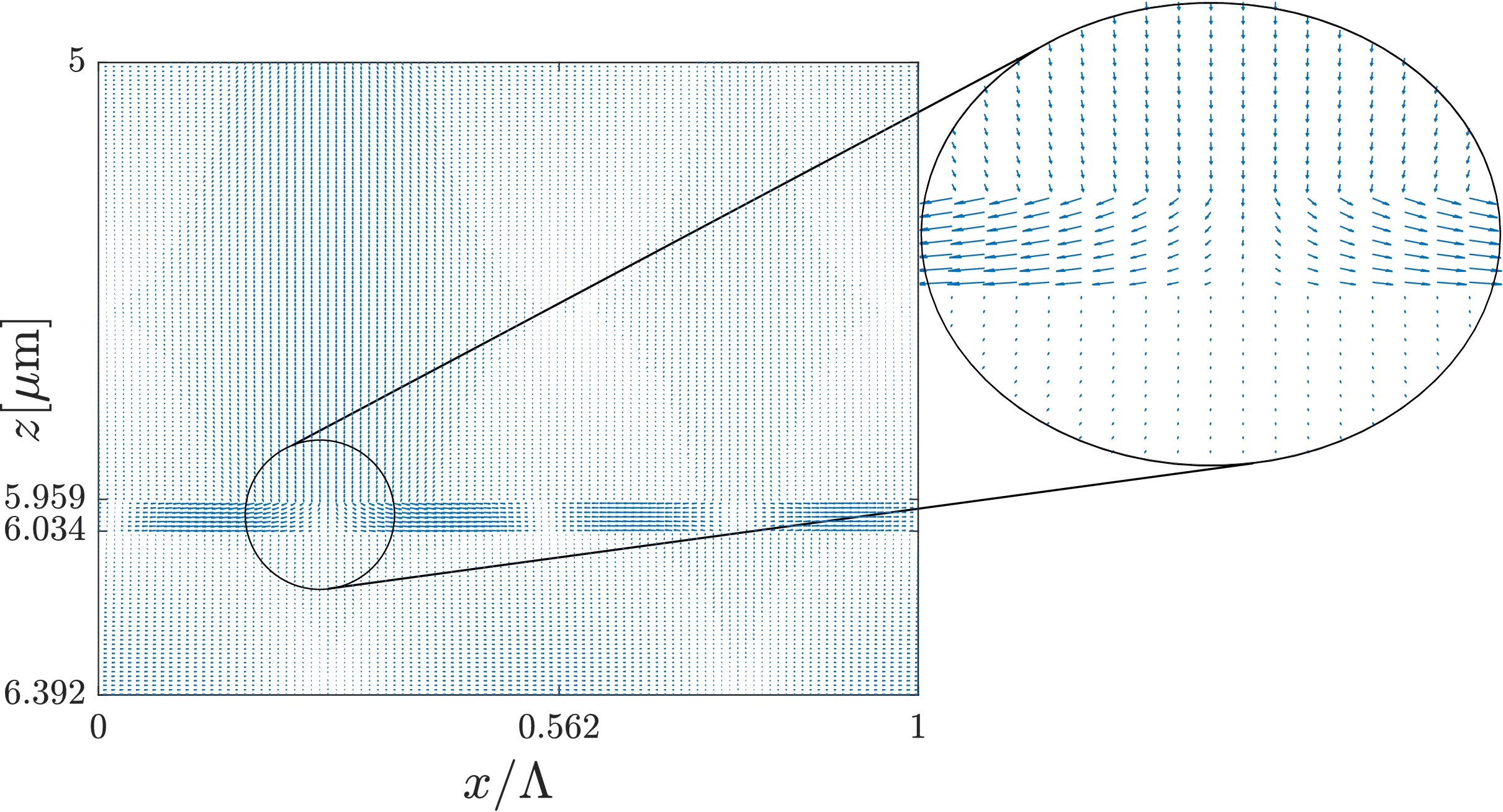}
\caption{{\label{fig:fig5}} The EM power flow map in the structure 2GLS-TM (Table \ref{tab:t2}). The quiver plot range and simulation wavelength are the same as in Fig.\ref{fig:fig7}.}
\end{figure*}

\begin{figure*}[htbp]
  \centering
  \includegraphics[width=14cm, height=7.53cm]{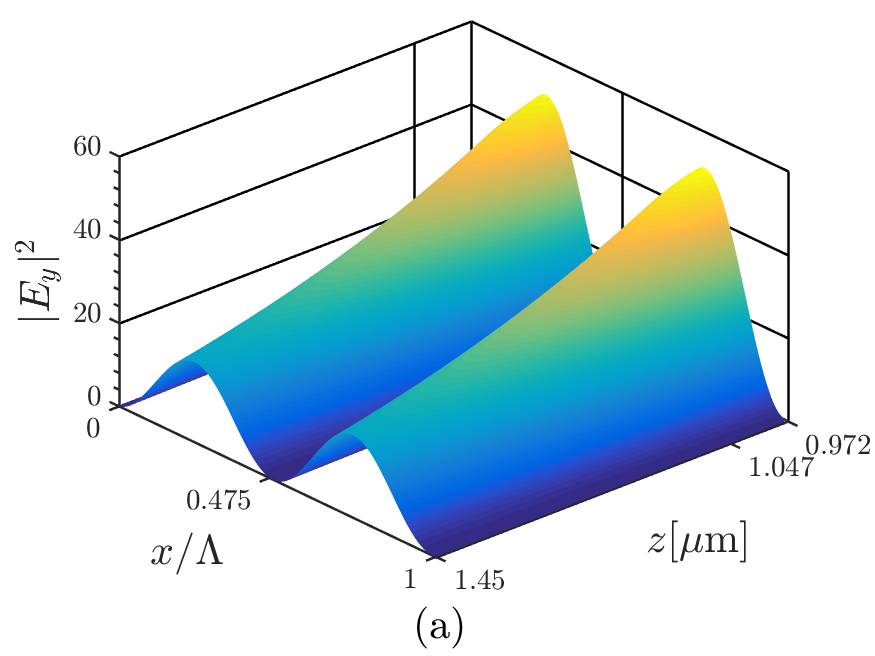}

  \includegraphics[width=14cm, height=7.53cm]{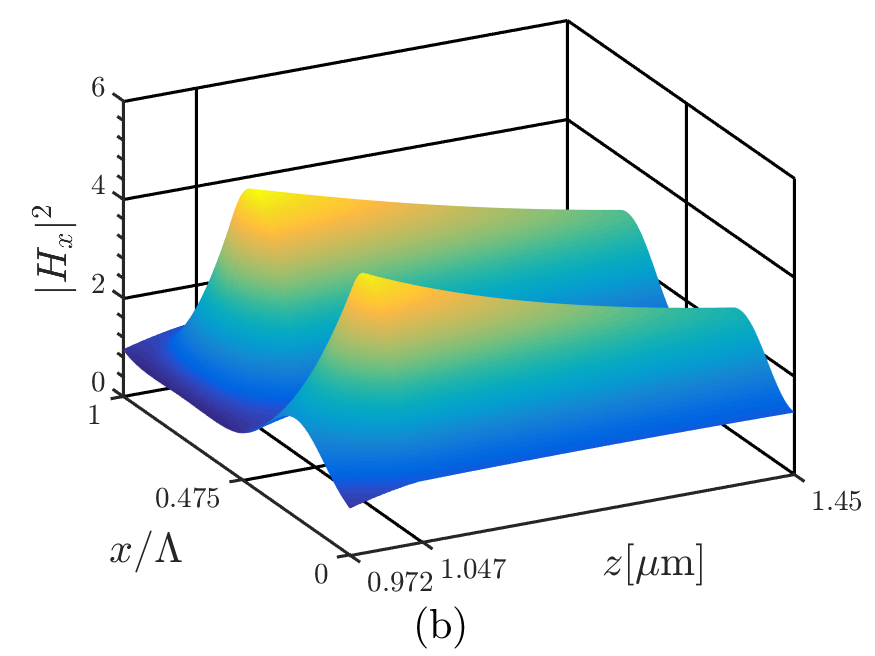}

  \includegraphics[width=14cm, height=7.53cm]{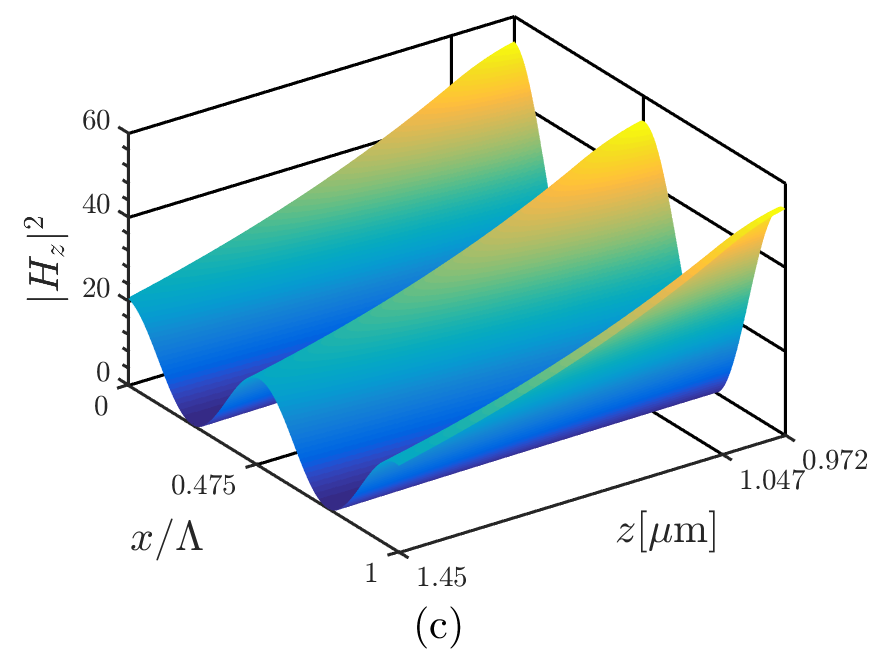}

\caption{\label{fig:fig6} The normalized to the input values amplitudes squared of the electric (a) and  magnetic (b), (c) fields' components at CDW $\lambda_0=4.415\hspace{0.02cm} \mu m$ in the structure 2GLS-TE (Table \ref{tab:t2}) vs. $0\leq x/\Lambda\leq 1$ and, as defined in the text, $0.972\hspace{0.02cm}\mu {\rm m}\leq z \leq 1.45\hspace{0.02cm}\mu {\rm m}$; $x/\Lambda=0.475$ is the inner grating lamella-groove interface derived line, $(0,1)\times(0.972\hspace{0.02cm}\mu {\rm m},1.047\hspace{0.02cm}\mu {\rm m})$ is the absorber cut within the ``microcavity''.}
\end{figure*}

\begin{figure*}[htbp]
  \centering
  \includegraphics[width=14cm, height=7.53cm]{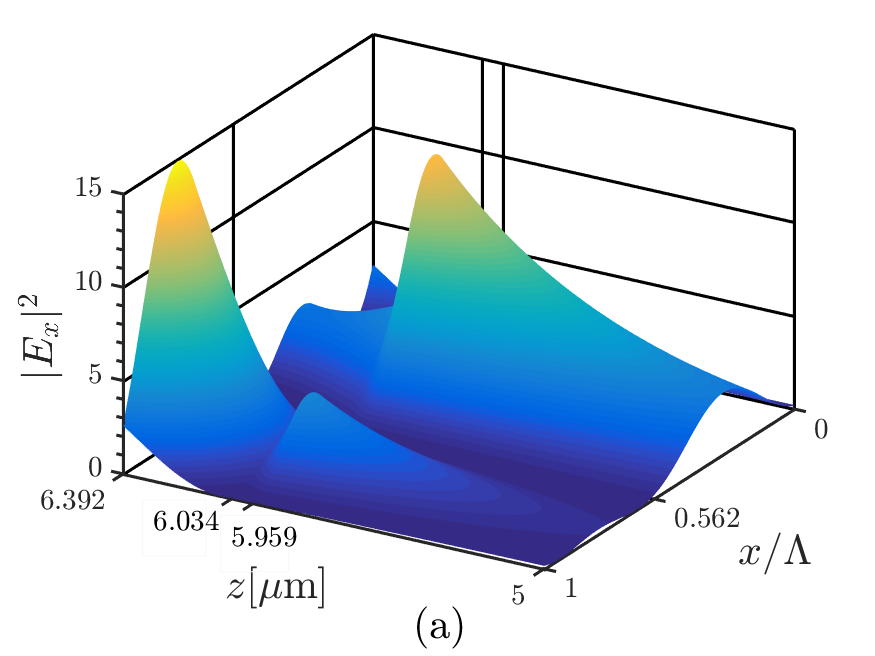}

  \includegraphics[width=14cm, height=7.53cm]{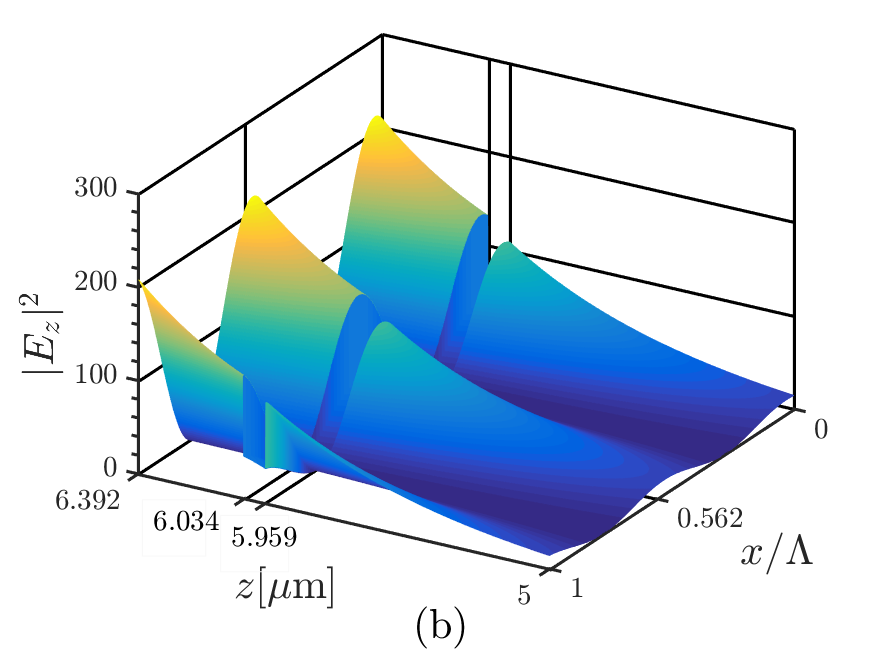}

  \includegraphics[width=14cm, height=7.53cm]{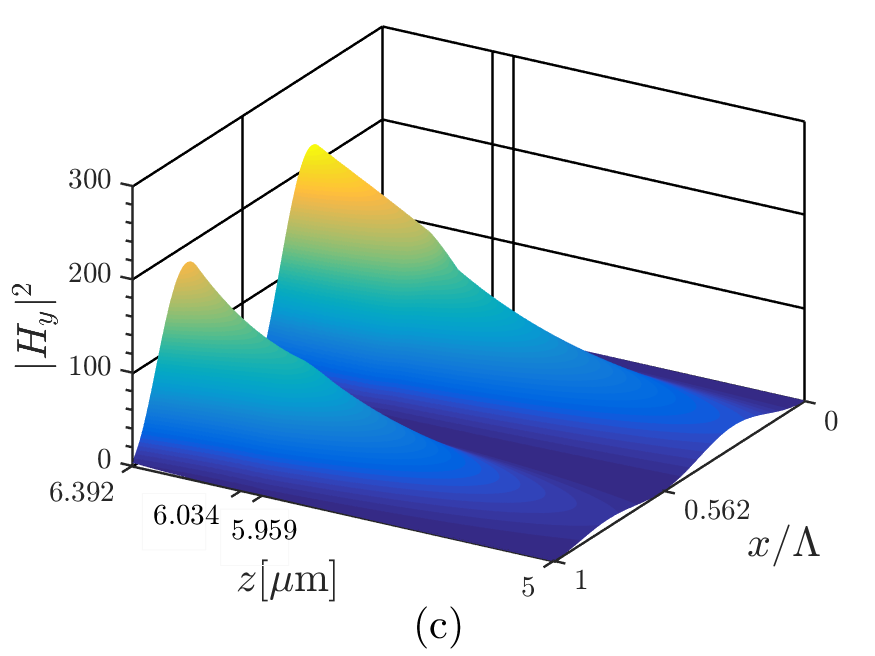}

\caption{{\label{fig:fig7}} The normalized to the input values amplitudes squared of the electric (a), (b) and  magnetic (c) fields' components at CDW $\lambda_0=4.415\hspace{0.02cm} \mu m$ in the structure 2GLS-TM (Table \ref{tab:t2}) vs. $0\leq x/\Lambda\leq 1$ and, as defined in the text, $5.\hspace{0.02cm}\mu {\rm m} \leq z \leq 6.392\hspace{0.02cm}\mu {\rm m}$; $x/\Lambda=0.562$ is the inner grating lamella-groove interface derived line, $(0,1)\times(5.959\hspace{0.02cm}\mu {\rm m},6.034\hspace{0.02cm}\mu {\rm m})$ is the absorber cut within the ``microcavity''.}
\end{figure*}

For TE- or TM-dual-GLS PDs, $E_y, H_x, H_z\neq 0$ or $H_y, E_x, E_z\neq 0$, respectively, and $\mathbf{S}\neq (0,0)$, so we plotted the surface graphs of the fields' components modules squared and the quiver maps of $\mathbf{S}$ vs. $x, z$ (Figs.\ref{fig:fig4}, \ref{fig:fig5}), where the plots' domains are rectangles in the $(x, z)$ plane. Due to $\Lambda$-periodicity of the physical quantities in $x$, any interval of the length $\Lambda$ might be suitable to eventually pick out the ``microcatity" $x$-range. Our initial choice is with $\left[0,\Lambda\right]$ overall, but it may further be revised depending on the structure. For the presentation convenience, of the above chosen $z$-range we visualize the ``microcatity" slices with $10\%$ of the front cavity in the 2GLS-TE PD, and with $20\%$ of the back cavity in the 2GLS-TM PD case, which contain the real lateral cavity-absorber interfaces and dummy vertical ones, reminiscent of the real gratings' lamella-groove interfaces. All nonzero $\left|E_i\right|^2$, $\left|H_k\right|^2$ of our scope are shown in Figs.\ref{fig:fig6} and \ref{fig:fig7}. Let us note the continuity of $\mathbf H$ as well as of $E_y$ and $E_x$, seen in Figs.\ref{fig:fig6}(a) and \ref{fig:fig7}(a), respectively, and the discontinuity of $E_z$ when crossing the cavity-absorber interfaces seen in Fig.\ref{fig:fig7}(b), which is fairly consistent with the EM boundary conditions.

%Fig 4

%Fig 5

%Fig 6

%Fig 7

\subsection{The EM power flow}
We will start with visualizing the EM power flow in the 2GLS-TE structure using the quiver plot of the Poynting vector $\mathbf{S}$ in Fig.\ref{fig:fig4}. It presents the EM power flow over the chosen "microcavity'' slice of that structure, clearly showing a major zero total circulation $\mathbf{S}$ vortex. It consists of two sub-vortices, namely, a split "microcavity'' boundary-adjoint and "microcavity''-interior vortices with the clockwise and counterclockwise $\mathbf{S}$ circulation, respectively, around respective vorticity centers, which locate $235$ nm above the absorber-cavity interface line $z=1.047\hspace{0.02cm} \mu$m. Yet, the sub-vortex adjoint to the boundary line $x = 0$ apparently has its vorticity center at a point located in the ``microcavity" adjacent left to the considered one, whereas the vorticity center of the vortex adjacent to the boundary line $x/\Lambda=1$ lies just inside the ``microcavity''. It appears that the distance between the vorticity centers are $0.5$ in the units of $\Lambda$, as may be expected from the $\Lambda$-periodicity of $\mathbf{|S|}$.

To symmetrize the vortex picture and thus to force the center of the boundary adjacent vortex to lie just on the vertical boundaries of the ``microcavity'' the latter can be rigidly shifted left to a distance of $0.0125\Lambda$. Anyway, as can be inferred from Fig.\ref{fig:fig4}, the major $\mathbf{S}$ vortex remains unconfined within the ``microcavity'' because of an appreciable lateral EM power flow between the ``microcavities''. In the interchange of the interior and boundary-adjoint sub-vortexes, just along mid-lines between the lamella-groove interface derived line and ``microcavity'' vertical boundary lines, the EM power flow is vertical, in positive and negative directions on the left and right to the interior sub-vortex, respectively. Vorticity of the EM power flow is a direct consequence of the continuity of $\mathbf{S}$, whereas a dense persistence of the $S_x$ component overall the absorber layer seems a key to the near $100\%$ efficiency of the 2GLS-TE structure.

In addition, as can be inferred from the zoomed inset in Fig.\ref{fig:fig4} there is a twist of $\mathbf{S}$ when its direction alternates to sustain its continuity, and these occurs just on the lamella-groove interface derived line $x/\Lambda=0.475$ (or $x/\Lambda=0.4875$ in the redefined ``microcavity''). Thus, the EM power flow strongly senses the presence of the gratings even remotely of them, in the smoothly layered ``microcavity'' part.

Next, we proceed to the 2GLS-TM structure. In this case the ``microcavity'' assisted absorption enhancement mechanism operates very differently from that of the 2GLS-TE structure considered above. While in the latter, we observe both vertical light flow traversing the absorber back and forth by reflection from one GLS to another, similarly to the case of the RCE absorber, and a persistent lateral light flow along the absorber layer, in the present case the light flow is $predominately$ lateral, see quiver plot in Fig.\ref{fig:fig3} and discussion below. In addition, in the TE case the $S_x$ continues to flow from the ``microcavity'' to the adjacent one, see quiver plot in Fig.\ref{fig:fig4}, whereas in the TM case it seems bouncing back and forth between the vertical absorber layer boundaries (inside the ``microcavity'' slice under consideration). In spite of such a difference the present mechanism also provides nearly $100\%$ efficiency.

In a detail, as seen from Fig.\ref{fig:fig5}, there are $\mathbf{S}$ density rarefaction vertical channel like areas in the absorber layer, three interior ones and another one which is split among the boundaries. As in the TE case the left and right boundary adjoint parts of the split "channel" seem asymmetric in the present ``microcavity'' choice, however, it is possible to symmetrize them provided that the ``microcavity'' is shifted by $0.031\Lambda$ to the left. The interior "channels" closest to the boundaries, see the left one in the inset of Fig.\ref{fig:fig5}, separated by a distance of 0.5 (in the units of $\Lambda$), seem as actually channeling light through the absorber, up through the left and down through the right "channels". Yet, such a vertical traversing the absorber is by far not dominate the lateral flow because of low relative thickness of those areas and rather small values of the $\mathbf{S}$ components in them as well. The main function of these real channels is to convert the mostly vertical flow of the incoming light into a predominantly lateral flow inside the absorber. The above noted mid-lines between the ``microcavity'' vertical boundary lines and the gratings' lamella-groove interface-derived lines cross the absorber just in the middle of those channels.

Concerning the other two "channels", upon the shift note above, the interior one moves to the redefined ``microcavity'' center and the boundary adjoint channel halves will start and end at the redefined ``microcavity'' vertical boundary lines. In the central "channel" there seen a very small $S_x$ component vanishing at the middle, as a result of the counter propagating lateral light flow which arrive from two real channels discussed previously. At the same time, the left and right parts of the boundary split "channel" serve as flow "sources" draining the light off the (redefined) ``microcavity'' to the adjacent ones.

When crossing the absorber's boundary lines $z=5.959\hspace{0.02cm} \mu$m and $z=6.034\hspace{0.02cm} \mu$m in the z-direction, $S_x$ exhibits jumps both in the magnitude and direction which is due to the discontinuity of the $E_z$ component as a function of $z$. Due to this peculiarity of the TM case, the absorber layer as a whole turns out to act as a laterally periodic absorbing waveguide. In spite of the lateral infiniteness of the structure, the periodicity dictated by the presence of the gratings, creates an evident lateral resonance the same in any ``microcavity'' so that the nearly $100\%$ absorbance is acquired within it.

In addition, in the back cavity, there occurs a twist of $\mathbf{S}$, which points to a positive z direction becomes more and more lateral until zeroing at a point lying on the lamella-groove interface derived line $x/\Lambda=0.562$ (or $x/\Lambda=0.531$ in the redefined ``microcavity'') and then returns to having the initial direction. Again, the EM power flow strongly senses the presence of the gratings even remotely of them, in the smoothly layered ``microcavity'' part.

\subsection{The EM Fields}
The surface plots of the relevant EM fields amplitudes squared in the 2GLS-TE structure are shown in Fig.\ref{fig:fig6}. The $|E_y|^2$ and $|H_z|^2$ surfaces, see in Figs.\ref{fig:fig6}a and c, respectively, show a similarity in attaining their maxima over the chosen ``microcavity'' slice just on the line $z=0.972\hspace{0.02cm} \mu m$, which is the boundary between the absorber and the back GLS, and comparability of their maxima. At the same time, those surfaces and $|H_x|^2$ in Fig.\ref{fig:fig6}c are different. First, $|H_x|^2$ maxima are an order of magnitude smaller than of those for the former ones. Second, while the $|E_y|^2$ and $|H_x|^2$ maxima, two for each, are remote from the vertical ``microcavity'' boundary lines, one of three apparent $|H_z|^2$ maxima lies left to the $x=0$ boundary. The boundary-adjoint $|H_z|^2$ patterns can be thought as a part of one surface due to the periodicity in $x$ but the left pattern is not symmetric with the right one. The symmetry can be restored by the above discussed ``microcavity'' shift done for the $\mathbf{S}$ quiver plot.
Inside the ``microcavity'' slice under consideration all the functions $|E_y|^2$, $|H_x|^2$ and $|H_z|^2$ exhibit a gradual decay when moving its argument point along the $z$-axis towards the front cavity boundary line (remind that we display only $10\%$ of it). On the contrary, those functions are decaying very sharply when moving their argument point along the $x$-axis towards some critical vertical lines. While in the case of the $|E_y|^2$ and $|H_x|^2$, the lines are ``microcavity'' vertical boundary ones and the line $x/\Lambda=0.4875$, which becomes the mid-line of the redefined ``microcavity'', in the case of the $|H_z|^2$ they are the mid-lines between the ``microcavity'' vertical boundary ones and the gratings' lamella-groove interface derived line. In addition, the maxima and minima of both  $|E_y|^2$ and $|H_z|^2$ complements each other respectively. The same is strictly valid also for the pair $|H_x|^2$ and $|H_z|^2$.

One can observe some behavioral trends making it possible to relate the fields amplitudes and $\mathbf{S}$ while this relationship may be not completely equivocal. It can be seen, e.g. that the decrease of $|H_x|$ ($H_x$ enters $S_z$) inside the absorber layer, which is seen in Fig.\ref{fig:fig6}b is reflected in the $\mathbf{S}$ direction change there, that is the sharpest change overall the vortex. In addition, the similar behavior of $|E_y|$ and $|H_z|$ seen in Figs.\ref{fig:fig6}a and \ref{fig:fig6}c, where the maximum regions complement the minimum ones is reflected in the opposite circularity of the interior and ``microcavity'' boundary-adjoint $\mathbf{S}$ sub-vortices.

There is a topological difference between the fields amplitude squared surfaces, and the EM power flows, for the TE and TM cases. Nevertheless, some regularities as concerns geometrical positions of critical lines and points and the appearance of ``microcavity'' boundary adjoint patterns of the EM fields $z$ components, remain similar with the circular replacement of $E \leftrightarrow H$. However, in the TM case there is a much more pronounced EM fields' vertical (along with lateral) confinement and electrical isolation of the ``microcavities''. In addition contrary to $|H_x|^2$ in the TE case, the maximum of $|E_x|^2$ in the TM case seen in Fig.\ref{fig:fig7}a do not lie on one horizontal line. This noticeable difference is supposedly due to the absence of a back cavity in the TE case.

Note that $E_z$ in a contrast with $H_z$ is not continuous when crossing the absorber-cavity interfaces as seen in Fig \ref{fig:fig7}b. That discontinuity of function $|E_z|^2$ does not affect its steep descent occurring when its argument move away the maxima locations along the $z$-axis towards the boundary line of the back cavity slice ($z=5\hspace{0.02cm} \mu$m) under consideration. In a detail, $|E_z|^2$ jumps down just in the absorber layer, then acquires a local maximum and next starts to continue the descent.
As in the TE case some qualitative relationship between the EM fields' amplitudes and $\mathbf{S}$ can be stated. For example, it can be seen that the decrease of the $H_y$ and $E_x$ components' amplitudes towards the absorber layer, would affect the $S_z$, which is indeed seen decreasing while steeply changing the $\mathbf{S}$ direction from being mostly vertical to predominantly lateral within the absorber layer. Moreover, due to the similar behavior of $|E_z|$ and $|H_y|$ that have maximum regions corresponding to minimum regions we can observe an opposite direction of the Poynting vector along the $x$-axis. The above noted "channels" in which $\mathbf{S}$ is vanishingly small, correlate well with the regions around the maxima of $|E_z|$ and minima $|H_y|$ as seen from a comparison of Fig.\ref{fig:fig5} with Figs.\ref{fig:fig7}b and \ref{fig:fig7}c, respectively.

\section{Design tolerances}
Last but not least is the fabrication tolerance issue which is of primary importance for practicality of the proposed PD structures. Therein, for manufacturing such devices needs only few layer deposition steps, the grating fabrication accuracy is most critical. Hitherto testing the sensitivity of the $\eta$ spectra of the designed one-GLS mirror RCE PDs to changes in the grating's etch depth and width around the design values, we found \cite{Zohar,OQE} ones with tolerances certainly practical from the current micro-fabrication status viewpoint.
%---------------------------------------------------
%Table 3
\begin{table}[!htb]
\begin{center}
\caption{\bf The performance tolerances of the dual-GLS PD structure for the TE polarization.$^a$}
\vspace*{2mm}
\label{tab:t3}
\begin{tabular}{|c|c|c|c|c|}
\hline
 \multicolumn{3}{|c|}{Etching errors} &   \multicolumn{2}{|c|}{2GLS-TE} \\
\hline
\rule[-1ex]{0pt}{3ex}$\Delta W/W$ & $\Delta t_{\rm{bg}}/t_{\rm{bg}}$ & $\Delta t_{\rm{fg}}/t_{\rm{fg}}$ & $\Delta \lambda_0, \rm{nm}$ & $\Delta \eta_{\max}, \%$\\
\hline
   $-$   & $-$   & $-$   &  $-$ 1.0   & $-$ 0.7    \\
\hline
  $-$    & $-$   & $+$   &  $-$ 1.0   & $-$ 1.2    \\
\hline
  $-$    & $+$   & $-$   &  $-$ 1.0   & $-$ 2.2    \\
\hline
 $-$     & $+$   & $+$   &  $-$ 1.0   & $-$ 0.6    \\
\hline
  $+$    & $-$   & $-$   &  $-$ 1.0   & $-$ 1.5    \\
\hline
  $+$    & $-$   & $+$   &  $-$ 1.0   & $-$ 1.6    \\
\hline
  $+$    & $+$   & $-$   &      0.0   & $-$ 0.1    \\
\hline
  $+$    & $+$   & $+$   &      0.0   & $-$ 0.1    \\
\hline
\end{tabular}
\end{center}
\begin{flushleft}$^a$Above $\Delta t_{\rm bg}/t_{\rm bg}$, $\Delta t_{\rm fg}/t_{\rm fg}$ and $\Delta W/W$ are the relative fabrication errors in the gratings grooves' depths and width, respectively, allowed for the etching; $\Delta \lambda_0$ is the shift of the wavelength of the $\eta(\lambda)$ maximum and $\Delta \eta_{\max}$ is the absolute change of the peak efficiency, which result from the fabrication errors. The single $+$ or $-$ sign abbreviates the relative error of $+10\%$ or $-10\%$.
\end{flushleft}
\end{table}
%Table 4
\begin{table}[!htb]
\begin{center}
\caption{\bf The performance tolerances of the dual-GLS PD structure for the TM polarization.$^a$}
\vspace*{2mm}
\label{tab:t4}
\begin{tabular}{|c|c|c|c|c|}
\hline
 \multicolumn{3}{|c|}{Etching errors} &   \multicolumn{2}{|c|}{2GLS-TM} \\
\hline
\rule[-1ex]{0pt}{3ex}$\Delta W/W$ & $\Delta t_{\rm{bg}}/t_{\rm{bg}}$ & $\Delta t_{\rm{fg}}/t_{\rm{fg}}$ & $\Delta \lambda_0, \rm{nm}$ & $\Delta \eta_{\max}, \%$ \\
\hline
   $-$   & $-$   & $-$   &  $-$ 2   & $-$ 13.8   \\
\hline
  $-$    & $-$   & $+$   &  $+$ 5   & $-$ 10.0   \\
\hline
  $-$    & $+$   & $-$   &  $-$ 1   & $-$ 14.4   \\
\hline
 $-$     & $+$   & $+$   &  $+$ 5   & $-$ 15.0   \\
\hline
  $+$    & $-$   & $-$   &  $-$ 3   & $-$ 13.6   \\
\hline
  $+$    & $-$   & $+$   &  $+$ 3   & $-$ 13.3   \\
\hline
  $+$    & $+$   & $-$   &  $-$ 2   & $-$ 14.7   \\
\hline
  $+$    & $+$   & $+$   &  $+$ 4   & $-$ 14.0   \\
\hline
\end{tabular}
\end{center}
\begin{flushleft}$^a$For the meaning and measures of the parameters here, see the descriptions in the text, and captions to Table \ref{tab:t3}.
\end{flushleft}
\end{table}

For the dual-GLS PDs under study, we performed similar inquiry, varying $t_{\rm{fg}}$, $t_{\rm{bg}}$ and $W$ around their optimal values in Table \ref{tab:t2} by $\pm 10\%$ of each. The variations are similar to adopting such minimal absolute errors in etching the grating grooves as $\pm 9.6\hspace{0.02cm}\rm nm$ of the depths and $\pm 70.1\hspace{0.02cm}\rm nm$ of the widths. The tolerance tests' results for the dual-GLS PD structures exhibiting the best peak efficiency upon TE- and TM-polarized light irradiation, are shown in Tables \ref{tab:t3} and \ref{tab:t4}, respectively.

Table \ref{tab:t3} covers the PD structure adapted to the TE polarization, and shows that under the allowed fabrication errors its sustain high performance, viz. exhibit concurrent small peak efficiency deterioration $\Delta \eta_{\max}<0$ and tiny peak-wavelength shift $\Delta \lambda_0$. In particular, a unique worst case emerges for the structure, such as with $\Delta \eta_{\max}= -2.2\%$, $\Delta \lambda_0=-1.0\hspace{0.02cm}\rm nm$ for 2GLS-TE. All other fabrication errors' combinations result in less than $2\%$ deterioration, with $\Delta\lambda_0$'s comparable to those in the respective worst case.
Table \ref{tab:t4} includes 2GLS-TM structure, for which the worst case is with: $\Delta\eta_{\max}= -15.0\%$, $\Delta \lambda_0=+5\hspace{0.02cm}\rm nm$. Here, $\Delta\eta_{\max}$'s (being an order of magnitude worse than for 2GLS-TE) and $\Delta \lambda_0$'s have relatively small scatter in the ranges from $-15\%$ to $-10 \%$ and $-3\rm nm$ to $ +5\rm nm$, respectively.
\section{Conclusions}
The concept of RCE PD responded the challenge of thinning the PD light absorbing layer while outperforming the efficiency of respective monolithic PD. However, for many relevant absorber materials, meeting that challenge with the RCE absorption requires so unrealistically many layer deposition steps that the fabrication faults become quite inevitable and the high efficiency at goal hardly attainable. Under this circumstance, we suggested to search PDs with the same absorber layer as in respective RCE PDs, but concurrently being much thinner and topmost performing.

In the above quest, we proposed a renewed PD structure which alike RCE PD contains the cavity-absorber arrangement but surrogates the DBR mirrors by two GLSs. By the design based on the in-house software, we showed the theoretical feasibility of such dual-GLS PDs of $\sim 100\%$ efficiency for the linearly polarized (TE or TM) MWIR radiation with thickness in the range $\sim 7.0\mu{\rm} m-8.5\mu\rm m$, see in Table \ref{tab:t2}. Furthermore, we tested the dual-GLS PDs' performance tolerance to the errors in gratings' groove etch, and revealed it to be superlatively high for TE-polarization targeted ones, see Table \ref{tab:t3}. TM polarization targeted dual-GLS PD, see in Table \ref{tab:t4}, tolerates the gratings' fabrication errors an order of magnitude weaker than its TE counterpart, though on the practical ground it is still fairly tolerant.

The EM fields' amplitudes and Poynting vector over a cavity-absorber area were visualized. As a result, it is inferred that the EM fields' confinement in the designed structure, which is a key to their upmost efficiency, is 2D one combining in-depth vertical resonant-cavity like confinement with the lateral microcavity like one set by the presence of gratings. The 2GLS-TE structure EM power flow showed vorticity like behavior as a direct consequence of the continuity of $\mathbf{S}$. The dense persistence of the $S_x$ component overall the absorber layer, and both vertical light flow traversing the absorber back and forth by reflection from one GLS to another, similarly to the case of the RCE absorber, and a persistent lateral light flow along the absorber layer seem the keys to the near $100\%$ efficiency of the 2GLS-TE structure. In the 2GLS-TM case, the ``microcavity'' assisted absorption enhancement mechanism was shown to operate very differently from that of the TE structure. The EM power flow is predominately lateral bouncing back and forth between the vertical absorber layer boundaries. Due to the discontinuity of the $E_z$ component as a function of $z$, the absorber layer as a whole turns out to act as a laterally periodic absorbing waveguide. Despite the lateral infiniteness of the structure, the periodicity dictated by the presence of the gratings creates an evident lateral resonance the same in any ``microcavity'' so that the nearly $100\%$ absorbance is acquired within it.  In both TE and TM cases, the EM power flow strongly senses the presence of the gratings even remotely of them as shown by the twist of $\mathbf{S}$ around the lamella-groove interface derived line in the smoothly layered ``microcavity'' part. Although the relationship may be not completely equivocal, there exist some behavioral trends relating the fields' amplitudes to $\mathbf{S}$.

Because of a high light-polarization sensitivity of the optical response from 1D gratings, and its lesser tolerance to the gratings dimensions' variation in the TM-polarization case, the thin PD designs like those carried out in this study may unequivocally benefit replacing the 1D gratings by 2D ones. PDs of the proposed type may have high potential in integrated multidetector arrays such as FPA, for the high capability of which highly efficient thin PD elements are vital.

% Bibliography
%\bibliography{MCT.bbl}

\end{document}